\documentclass[twocolumn,preprintnumbers,aps,showpacs,superscriptaddress,prx]{revtex4-1}

\usepackage{bm}
\usepackage[pdftex]{graphicx}
\usepackage{latexsym}
\usepackage{epsfig}
\usepackage{amsmath,amssymb}
\usepackage{amsfonts}
\usepackage{palatino}
\usepackage[latin1]{inputenc}
\usepackage{epsf}
\usepackage{latexsym}
\usepackage{calc}
\usepackage{color}
\usepackage{shadow}
\usepackage{epsfig}
\usepackage{float}
\usepackage[all]{xy}
\usepackage[pdftex, pdfstartview = {FitH}]{hyperref}
\usepackage{csquotes}
\usepackage{afterpage}

\newcommand{\ben}{\begin{equation}}
\newcommand{\een}{\end{equation}}
\newcommand{\bea}{\begin{eqnarray}}
\newcommand{\eea}{\end{eqnarray}}

\newcommand{\nn}{\nonumber}

\def\k{{\bf k}}
\def\q{{\bf q}}

\begin{document}

\title{Coulomb and tunneling coupled trilayer systems at zero magnetic field}

\date{\today}

\author{D. Miravet}
\email{dmiravet@gmail.com}
\affiliation{Centro At\'omico Bariloche, 8400 S. C. de Bariloche, R\'io Negro, Argentina}
\author{C. R. Proetto}
\email{proetto@cab.cnea.gov.ar}
\affiliation{Centro At\'omico Bariloche and Instituto Balseiro, 8400 S. C. de Bariloche, R\'io Negro, Argentina}
\author{P. G. Bolcatto}
\affiliation{Instituto de F\'isica del Litoral (CONICET-UNL) and Facultad de Humanidades y Ciencias (Universidad Nacional del Litoral), G\"uemes 3450, 3000 Santa Fe, Argentina.}

\begin{abstract}
The ground-state electronic configuration of three coupled bidimensional electron gases has been determined using a
variational Hartree-Fock approach, at zero magnetic field. The layers are Coulomb coupled, and tunneling is present between
neighboring layers. In the limit of small separation between layers, the tunneling becomes the dominant energy
contribution, while for large distance between layers the physics is driven by the Hartree electrostatic energy.
Transition from tunneling to hartree dominated physics is shifted towards larger layer separation values as the total 
bidimensional density of the trilayers decreases. The inter-layer exchange stabilizes a ``balanced'' 
configuration, where the three layers are approximately equally occupied; most of the experiments are performed in the vicinity of this balanced configuration. 
Several ground-state configurations are consequence of a delicate interplay between tunneling and inter-subband exchange.
\end{abstract}

\maketitle

\section{Introduction}
\label{introduction}

Single-layer quasi two-dimensional electron gases (2DEG) such as those formed at the interface between two
dissimilar semiconductors can be routinely driven to a many-body interaction regime
by application of a strong magnetic field of several Teslas perpendicular to the 2DEG layer~\cite{Bastard}. By increasing 
the magnetic field, the 2DEG first enters in the Integer Quantum Hall regime, and then in the Fractional Quantum Hall
regime~\cite{DSP97}. 

Multilayer coupled 2DEG's like trilayer systems offer much more possibilities for the 
theoretical~\cite{HMcD96,HP00,Ye05} and
experimental~\cite{SESS92,LSJYS96,SSS98,WMGRBP09,GWRBP09,WMGRBP10,SBGLSBP14} search of the involved physics,
even at zero magnetic field. 
New single-particle effects like layer (site) energies, tunneling coupling between layers, and many-body effects as the 
inter-layer Coulomb coupling should be included,
with the two later effects being strongly dependent on the separation between layers.

\begin{figure}[h]
\begin{center}
\includegraphics[width=8.6cm]{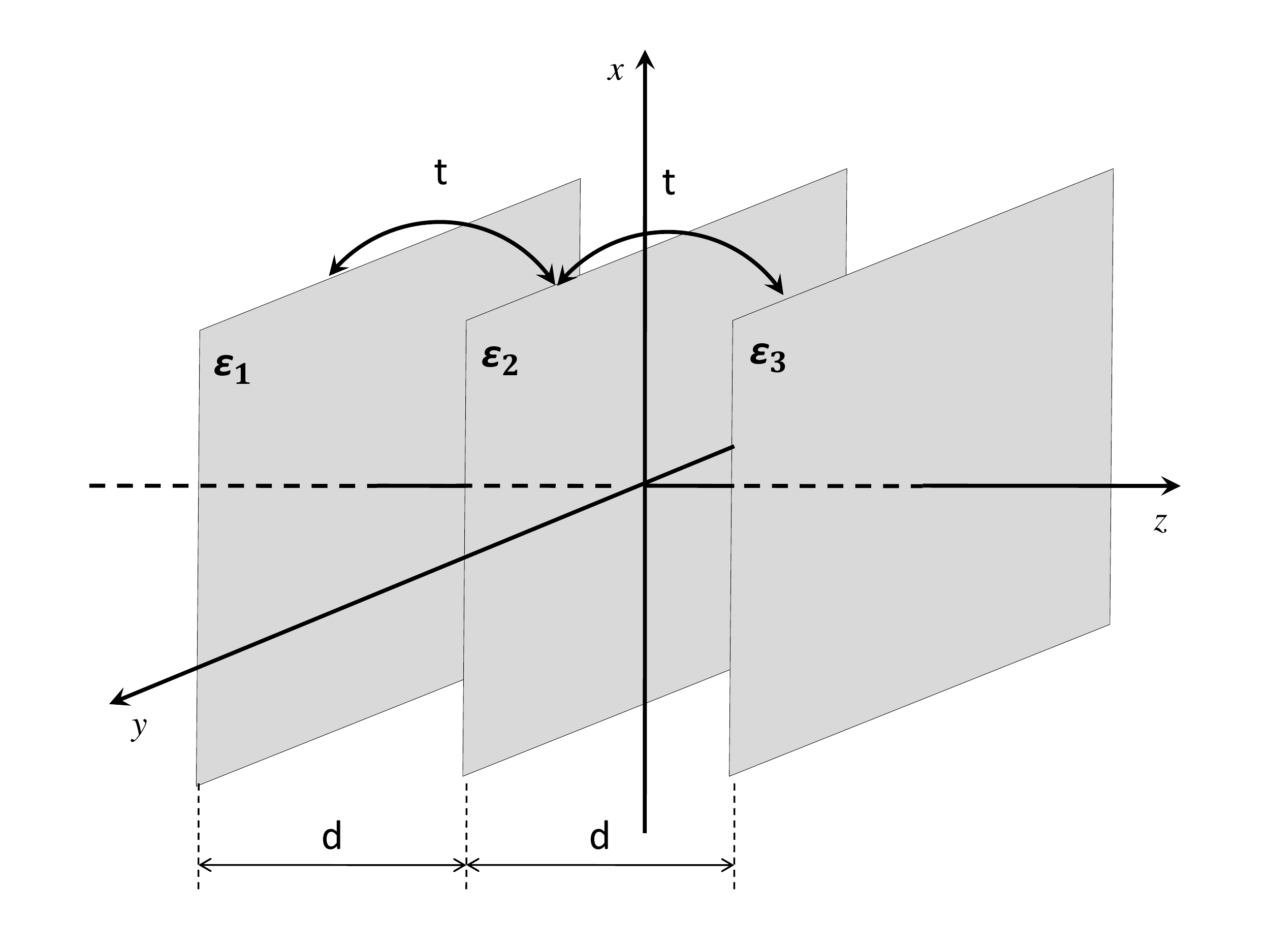}
\caption{\label{Fig1}Schematic view of the three layer system. Each layer of area $A$ has a two-dimensional electronic number
 density given by $n_1$, $n_2$, and $n_3$. Charge neutrality is provided by two positively charged layers located at 
 $z=\pm \: h$ (not shown), with $h \gg d$. The layers are coupled by the hopping $t$ and the Coulomb interaction.
 $\varepsilon_1$, $\varepsilon_2$, and $\varepsilon_3$ are the site or layer energies.}
\end{center}
\end{figure}

We provide in this work an exhaustive theoretical study of trilayer systems at zero-magnetic field, within the framework
of a Hartree-Fock mean field approximation, already used for bilayer systems~\cite{HHDV00}, and also for a
simplified version of the trilayer system considered here~\cite{HP00}. The model, schematically 
illustrated in Fig.~\ref{Fig1}, includes intra-layer and inter-layer (hopping) kinetic energies
(not considered in Ref.~[\onlinecite{HP00}]), site (layer) energies (also not considered in Ref.~[\onlinecite{HP00}]),
intra-layer and inter-layer exchange energies, and the usually dominant Hartree electrostatic energy. An additional
external parameter is the total density of the system: we will see that as the density decreases, the trilayer
system is driven to a single-particle tunneling dominated regime.

By doing a numerical minimization of the total energy of the system over all the internal variables, we have
found a zero-temperature extremely rich phase-diagram for the possible ground-state configurations. These full numerical
results may be qualitatively understood as resulting from the diverse scaling properties of the different
energy contributions to the total energy of the system, with respect to the external variables like the 
total density or the distance between layers.
Interestingly, we have found that at zero magnetic field and for close enough layers the physics of the trilayer system becomes dominated by
single-particle effects {\em both} in the high-density limit (dominated by the intra-layer kinetic energy) and
in the low-density limit (dominated by the hopping or tunneling and site energies). The interaction
dominated regime is then restricted to intermediate densities (exchange) and large separations between layers (Hartree).
The wide ground-state exploration in the available parameter space (distance between layers, total electronic density,
tunneling coupling) provided in this work may serve as a qualitative guide for the design of experimental
trilayer systems.

The rest of the work is organized in the following way: in Section II we explain the model and introduce the 
variational Hartree-Fock method we use to obtain the expression for total energy of the system. Analytical and
numerical ground-states resulting from the minimization of the system total energy are presented in Section III. 
Section IV is devoted to the conclusions. In the two appendices we discuss some limits and the physics beyond
the inter-layer exchange energy contribution (Appendix A), and we explain how to model the hopping parameter dependence
with the layer separation (Appendix B).

\section{Model}
Typically, a semiconductor trilayer system is realized experimentally by confining three {\it GaAs} quantum wells among 
{\it AlAs} or {\it Al$_x$Ga$_{1-x}$As} barriers of variable width and height,
which gives a control on the tunneling coupling $t$ among layers. Usually, the central $GaAs$ quantum well 
is designed with a larger width than the two
side-wells with the purpose of populate the central well, which tends to be less 
populated than the two side-wells\cite{P99,SSJMcD99}. In our model of Fig.~1, we can
simulate this feature by imposing, for instance, that $\varepsilon_2 < \varepsilon_1=\varepsilon_3$.

The model for our Coulomb and tunneling coupled trilayer system is represented schematically in Fig.~1, and defined
through the corresponding Hamiltonian in Eq.~(\ref{Hamiltonian}). It consists of
three strictly two-dimensional metallic layers, at a distance $d$ between them. Two other layers, located at
$z = \pm \: h$ along the $z$-axis ($h \gg d$) provide the compensating positive charge densities. Neighboring 
metallic layers
are coupled through the tunneling term $t$ (hopping), while all layers are Coulomb coupled through the inter-layer
Coulomb interaction.

In detail, the Hamiltonian of our model is as follows,
\begin{eqnarray}
 \hat{H} &=& \sum_{j \k \sigma} \varepsilon_{j k} c^{\dagger}_{j \k \sigma} c_{j \k \sigma}
 -t \sum_{j_1 j_2 \k \sigma} (c^{\dagger}_{j_1 \k \sigma} c_{j_2 \k \sigma} + c^{\dagger}_{j_2 \k \sigma} c_{j_1 \k \sigma}) \nonumber \\
 &-& \sum_{j \k \sigma} \sum_{m} V_{jm}(0) \, p_m \, c^{\dagger}_{j \k \sigma} c_{j \k \sigma} 
 + \frac{A}{2}\sum_{mn} V_{mn}(0) \, p_m p_n \nn \\
 &+& \sum_{\substack{j_1 \k_1 \sigma_1 \\ j_2 \k_2 \sigma_2 \\ \q}} \frac{V_{j_1j_2}(q)}{2A}
 c^{\dagger}_{j_1 \k_1+\q \sigma_1} c^{\dagger}_{j_2 \k_2-\q \sigma_2}c_{j_2 \k_2 \sigma_2} c_{j_1 \k_1 \sigma_1} \; ,
\label{Hamiltonian}
\end{eqnarray}
with $A$ denoting the area of each layer.
Here, $c^{\dagger}_{j \k \sigma}$ ($c_{j \k \sigma}$) is a creation (annihilation) operator for an electron
in layer $j$ ($j=1,2,3$), with two-dimensional momentum $\k$ and spin $\sigma$ ($\uparrow$ or $\downarrow$).
Each one of the five terms in the Hamiltonian represents a different physical contribution to the total energy.
The first corresponds to the sum of the layer and kinetic energy of electrons in each layer,
with $\varepsilon_{j k} = \varepsilon_j + \hbar^2 k^2 / 2 m^*$ ($m^*$ being the electron effective mass of the
well-acting semiconductor). The second term represents the quantum tunneling of electrons among different layers,
with a hopping amplitude parametrized by $t$ ($>0$); $j_1 \ne j_2$, and the sum is restricted to neighboring layers. 
The last three Coulomb-related terms represent the attractive ion (positive layer) - electron interaction,
the ion (positive layer) - ion (positive layer) interaction, and the repulsive electron - electron interaction,
respectively.
$p_m$ ($m=L,R$) denotes the uniform density of the two layers located at left and right of the three central
layers. The system fulfills a global neutrality condition, $N=n_1+n_2+n_3=p_L+p_R$, where $N$ represents the total
two-dimensional number density, and $n_j$ ($j=1,2,3$) are the number densities of each metallic layer.
Finally, $V_{ij}(q)=\frac{2 \pi e^2}{\epsilon q} e^{-qd_{ij}}$, with $d_{ij}=0, d, h$; $\epsilon$ is the dielectric
constant of the well-acting semiconductor ($\sim$ 12.5 for {\it GaAs}). For $t=0$ and 
$\varepsilon_1=\varepsilon_2=\varepsilon_3$, the model reduces to the one studied previously in Ref.~[\onlinecite{HP00}].
The layer energies are changed at will in real samples by the application of back and front gates, which in our
case are represented by the two positively charged layers at $\pm \: h$. In this work we consider only  the symmetric configurations, i.e. $p_L=p_R$,
and $\varepsilon_1=\varepsilon_3 > \varepsilon_2$.

The presence of the last electron-electron interaction term in Eq.~(\ref{Hamiltonian}) implies that no exact solution
is available for the model, and forces us to attempt its approximate solution. 
For this, we will employ a Hartree-Fock variational approximation, widely used for the bilayer case,
either at zero~\cite{HHDV00} or with magnetic field~\cite{B90,JSSSMcD98,JMcD00}, 
and also used in the previous trilayer zero-tunneling and zero-site energy study~\cite{HP00}. But before that,
we find convenient to perform an exact transformation of the Hamiltonian, by defining the following operators:
\begin{eqnarray}
 a^{\dagger}_{\k \sigma} &=& \cos^2 \left( \frac{\theta}{2} \right) e^{-i\phi} \: c^{\dagger}_{1 \k \sigma} +
 \sqrt{2} \sin\left(\frac{\theta}{2}\right)\cos\left(\frac{\theta}{2}\right)c^{\dagger}_{2 \k \sigma} \nn \\
 &+& \: \sin^2\left(\frac{\theta}{2}\right) e^{i\phi} \: c^{\dagger}_{3 \k \sigma} \; , \nonumber \\
 b^{\dagger}_{\k \sigma} &=& -\frac{\sin \theta}{\sqrt{2}} e^{-i\phi} \: c^{\dagger}_{1 \k \sigma} +
 \cos \theta \: c^{\dagger}_{2 \k \sigma} 
 + \frac{\sin\theta}{\sqrt{2}} e^{i\phi} \: c^{\dagger}_{3 \k \sigma} \; , \nonumber \\
 c^{\dagger}_{\k \sigma} &=& \sin^2 \left( \frac{\theta}{2} \right) e^{-i\phi} \: c^{\dagger}_{1 \k \sigma} -
 \sqrt{2} \sin\left(\frac{\theta}{2}\right)\cos\left(\frac{\theta}{2}\right)c^{\dagger}_{2 \k \sigma} \nn \\
 &+& \: \cos^2\left(\frac{\theta}{2}\right) e^{i\phi} \: c^{\dagger}_{3 \k \sigma} \; , \nonumber \\ 
 \label{ct}
\end{eqnarray}
with $0 \le \theta \le \pi$, and $0 \le \phi < 2\pi$.
The new operators satisfy fermion anticommutation relations
$\{\alpha_{\k \sigma},\alpha^{\dagger}_{\k' s'} \}= \delta_{\k,\k'}\delta_{\sigma,\sigma'}$, while all others
anticommutators vanishes. The transformation to the ``subband'' basis $a^{\dagger}_{\k \sigma}$, $b^{\dagger}_{\k \sigma}$, 
$c^{\dagger}_{\k \sigma}$ from the ``layer'' basis 
$c^{\dagger}_{1\k \sigma}$, $c^{\dagger}_{2\k \sigma}$, $c^{\dagger}_{3\k \sigma}$ is defined by the two angles $\theta$ 
and $\phi$.
For $\theta=\phi=0$, both basis are the same. For $\theta=\pi/2, \phi=0$,
$a^{\dagger}_{\k \sigma}$, $b^{\dagger}_{\k \sigma}$, and $c^{\dagger}_{\k \sigma}$ are creation operators for electrons 
in the 
ground- (symmetric, no nodes), first-excited (antisymmetric, one node), and second-excited (symmetric, two nodes)
subband states of the trilayer system.
The canonical transformation of Eq.~(\ref{ct}) may be considered as a general rotation in a 3-dimensional pseudo-spin
Hilbert space of a pseudo-spin spinor pointing in the 
$\hat{{\bf n}} =(\sin\theta \cos\phi, \sin\theta \sin\phi, \cos\phi)$ direction, with the layer index $j$ playing the role
of the pseudo-spin components. In the ground-state, the direction of $\hat{{\bf n}}$ 
is determined by the total energy minimum.
The coefficients in the $3 \times 3$ transformation matrix in Eq.~(\ref{ct}) are obtained from the eigenstates of 
the $3 \times 3$ matrix 
$\hat{{\bf n}}(\phi,\theta) \cdot {\bf S} = 
n_x(\phi,\theta) S_x + n_y(\phi,\theta) S_y + n_z(\phi,\theta) S_z$, with
$S_x$, $S_y$ and $S_z$ being the components of the $3 \times 3$ angular momentum matrices corresponding to {\it unit
pseudo-spin} (or angular momentum)~\cite{G96}. Write in the $a,b,c$ basis, the hopping term in Eq.~(\ref{Hamiltonian}) 
becomes
diagonal for the choice $\phi=0, \theta=\pi/2$, but our variational ansatz given by Eq.~(\ref{ansatz}) below is more
flexible and allows $\phi$ and $\theta$ to take any value within their permissible range. On the other side,
the transformation of Eq.~(\ref{ct}) is not the more general one, considering that the angles $\phi$ and $\theta$ may
be, in principle, $\sigma$ and $\k$ dependent. For simplicity, we have not included these dependences in our calculations. 

After expressing the Hamiltonian in Eq.~(\ref{Hamiltonian}) in term of the subband operators, we have taken the 
expectation value of the transformed Hamiltonian with the following Hartree-Fock variational ansatz for the 
ground state-vector~\cite{canted},
\begin{widetext}
\begin{equation}
 \vert \Psi_0 \rangle = 
 \prod\limits_{\k_6}^{k_6 \le k_{c\downarrow}}c^{\dagger}_{\k_6\downarrow}
 \prod\limits_{\k_5}^{k_5 \le k_{c\uparrow}}c^{\dagger}_{\k_5\uparrow}
 \prod\limits_{\k_4}^{k_4 \le k_{b\downarrow}}b^{\dagger}_{\k_4\downarrow}
 \prod\limits_{\k_3}^{k_3 \le k_{b\uparrow}}b^{\dagger}_{\k_3\uparrow}
 \prod\limits_{\k_2}^{k_2 \le k_{a\downarrow}}a^{\dagger}_{\k_2\downarrow}
 \prod\limits_{\k_1}^{k_1 \le k_{a\uparrow}}a^{\dagger}_{\k_1\uparrow} \vert 0 \rangle \; .
 \label{ansatz}
\end{equation}
\end{widetext}
Here, $k_{a\sigma}$, $k_{b\sigma}$, and $k_{c\sigma}$ are the Fermi wavectors for electrons with spin $\sigma$ in subbands $a$, $b$, and $c$,
respectively. If any of the six $k_{\alpha \sigma} = 0$, this means that the corresponding subband is empty. 
We obtain, after a lengthy calculation
\begin{eqnarray}
 \frac{\langle \Psi_0 \vert \hat{H} \vert \Psi_0 \rangle}{\text{Ry}^*\:N} &=& \frac{2}{r_s^2}\big[
 E_0 (\eta_{a\uparrow},\eta_{a\downarrow},\eta_{b\uparrow},\eta_{b\downarrow},\eta_{c\uparrow},\eta_{c\downarrow},
  \theta, \phi) \big] \; , \nn \\
 &=& \frac{2}{r_s^2} 
 \big[ E_0^{\,\text K} + E_0^{\,\text T} + E_0^{\,\text H} + E_0^{\,\text{X-intra}} + E_0^{\,\text{X-inter}}\big] \; , \nn \\
 \label{HF}
\end{eqnarray}
where
\begin{eqnarray}
 E_0^{\,\text K} &=&  \sum_{\alpha \sigma} \eta_{\alpha \sigma}^2   
                          + \frac{r_s^2}{2}\left\{ \varepsilon_2^* \left[ (\eta_{a}+\eta_{c}) \frac{\sin^2\theta}{2} + 
                                                         \eta_{b} \cos^2 \theta \right] \right. \nn \\
                         &+& \left. \varepsilon_1^* \left[ \eta_a \cos^4\left(\frac{\theta}{2}\right) + 
                                                           \eta_{b} \frac{\sin^2\theta}{2} +
                                                           \eta_c \sin^4\left(\frac{\theta}{2}\right)  
                                                         \right] \right. \nn \\
                         &+& \left. \varepsilon_3^* \left[ \eta_a  \sin^4\left(\frac{\theta}{2}\right) + 
                                                           \eta_{b} \frac{\sin^2\theta}{2}  +
                                                           \eta_c \cos^4\left(\frac{\theta}{2}\right)  
                                                         \right] \right\} \; , \nn \\                               
\label{K}
\end{eqnarray}
\begin{equation}
 E_0^{\,\text T} = - \frac{r_s^2}{\sqrt{2}}\:t^*\:(\eta_a-\eta_c)\:\sin\theta\:\cos\phi \; ,
\end{equation}
\begin{eqnarray}
 E_0^{\,\text H} &=& -2d^*\bigg\{\eta_b(\eta_a+\eta_c)+2\eta_a\eta_c+ \nn \\
 &+&\frac{\sin^2\theta}{2}\left[(\eta_a-\eta_c)^2+2\eta_b^2-\eta_b(\eta_a+\eta_c)\right] \nn \\
 &-&\frac{\sin^4\theta}{8}(\eta_a-2\eta_b+\eta_c)^2 \bigg\} \; ,
 \label{Hartree}
\end{eqnarray}
\begin{equation}
 E_0^{\,\text{X-intra}} = -\frac{8\,r_s}{3\pi}\sum_{\alpha \sigma} \eta_{\alpha \sigma}^{3/2} \; ,
 \label{X-intra}
\end{equation}
and
\begin{eqnarray}
 E_0^{\,\text{X-inter}} &=& r_s \sin^2\theta\sum_{\sigma}\int\limits_0^{\infty} dq \left(1-e^{-2d^*q/r_s}\right) \nn \\
                              &\times& \left(I_{aa\sigma}+2I_{bb\sigma}+I_{cc\sigma}-2I_{ab\sigma}-2I_{cb\sigma} \right) \nn \\
                              &-& \frac{r_s \sin^4\theta}{8}\sum_{\sigma}\int\limits_0^{\infty} dq 
                              \left( 3-4e^{-2d^*q/r_s}+e^{-4d^*q/r_s} \right) \nn \\
                              &\times& \left( I_{aa\sigma}+4I_{bb\sigma}+I_{cc\sigma}+2I_{ac\sigma}-4I_{ab\sigma}-4I_{cb\sigma} \right) \; .\nn \\ 
\label{X-inter}
\end{eqnarray}
Here, $r_s = 1/(a_0^* \sqrt{\pi N})$ is the dimensionless two-dimensional density parameter, 
$a_0^*=\epsilon \hbar^2/e^2 m^*$ is the effective Bohr radius of the well-acting semiconductor, 
and $\text{Ry}^* = m^* e^4 / (2 \epsilon^2 \hbar^2)$ is the effective Rydberg. For 
{\it GaAs} as well-acting material, $m^* \simeq 0.067 m_0$ with $m_0$ denoting the bare electron mass,
and $\epsilon \simeq 12.5$, resulting in $\text{Ry}^* \simeq 5.83$ meV, and $a_0^* \simeq 98.7$ \AA.
Also, $\eta_{\alpha}=\sum_{\sigma}\eta_{\alpha \sigma}$, 
$\eta_{\alpha \sigma}=\sum_{\k}\langle \alpha^{\dagger}_{\k \sigma}\alpha_{\k \sigma}\rangle/AN$ are the total and spin-discriminated
subband occupation factors,  
$d^*=d/a_0^*$ is the distance from the central layer to the two lateral layers 
(in units of the effective Bohr radius $a_0^*$), $t^*=t/\text{Ry}^*$, and $\varepsilon_i^*=\varepsilon_i/\text{Ry}^*$. 
The expression for 
the quantities $I_{\alpha \beta \sigma}(q)$ is given in the Appendix A.
These exchange integrals correspond to the overlap area between two Fermi circles, with radius $k_{\alpha,\sigma}$ and
$k_{\beta,\sigma}$, with the circles centers separated by a distance $q$. The overlap area is maximum for $q=0$,
and decreases as $q$ increases; when $q \geq k_{\alpha,\sigma} +k_{\alpha,\sigma}$ the superposition area 
(and the associated exchange integral) becomes zero. 
All energies in Eq.~(\ref{HF}) are given in units of
$N \; \text{Ry}^*$, i.e., in units of energy per unit area.
As defined above, all $0 \le \eta_{\alpha \sigma} \le 1$, and $\sum_{\alpha \sigma} \eta_{\alpha \sigma} = 1$.

$E_0^{\text K}$ corresponds to the sum of the intra-layer kinetic and layer energy contributions. For the
case $\varepsilon_1^*=\varepsilon_2^*=\varepsilon_3^*$, the layer energy contribution reduces to an uninteresting constant term, which only
depends on the total electron density~\cite{HP00}. $E_0^{\text T}$ is the tunneling or inter-layer kinetic energy, and is the only term
where the angle $\phi$ appears. $E_0^{\text H}$ is the Hartree electrostatic energy, and ${E}_0^{\,\text{X-intra}}$ 
and ${E}_0^{\,\text{X-inter}}$ are the intra-subband and inter-subband exchange-energy contributions, respectively.
For a given set of external parameters ($r_s$, $d^*$, $t^*$, $\varepsilon_1^*$, $\varepsilon_2^*$, $\varepsilon_3^*$), the 
ground-state energy $E_0$ depends on eight variational parameters:
$\eta_{a\uparrow}, \eta_{a\downarrow}, \eta_{b\uparrow}, \eta_{b\downarrow}, \eta_{c\uparrow}, \eta_{c\downarrow},
  \theta, \phi$.
However, the neutrality condition $\eta_a + \eta_b + \eta_c = 1$ allows us to eliminate 
one of the six subband occupation factors. Regarding the hopping term, since
$\sin\theta \ge 0$ for $0 \le \theta \le \pi$, for having $E_0^{\text T} \le 0$, the condition is that 
$(\eta_a-\eta_c) \cos\phi \ge 0$.
As noted above, the total energy in Eq.~(\ref{HF}) is invariant under the exchange of the subband
labels ``$a$'' and ``$c$''. And as the angle $\phi$ only enters through the hopping term $E_0^{\text T}$, its value is
just determined by the sign of $\eta_a - \eta_c$. For example,
by assuming that we restrict ourselves to the configurations with $\eta_a \ge \eta_c$, the optimum value
for $\phi$ is $\phi=0$. Having assumed instead that $\eta_a \le \eta_c$ for all possible configurations, the optimum value for $\phi$ will be $\phi=\pi$.
Both choices are equivalent, and we have adopted here the first one: $\eta_a \ge \eta_c, \phi = 0$.
Under these constraints, $E_0$ has been minimized numerically with respect to the remaining six variational parameters:
five occupation factors and the layer mixing angle $\theta$. The corresponding results are presented in Figs.~2-7.

The scaling with $r_s$ and $d^*$ of the different energy contributions deserves some discussion. In the low-density
limit, $r_s$ grows since it is inversely proportional to the square root of the density and the physics 
is dominated by the 
tunneling and layer contributions, which are single-particle
effects. On the other side, in the high-density limit $r_s$ is small, and the main contributions to the total energy
comes from the intra-layer kinetic and Hartree terms. For large enough $d^*$, on the other side,
the system always becomes dominated by the Hartree term, which favors an effective bilayer configuration with the central well
empty, and with electrons distributed equally in the two side layers~\cite{P99,SSJMcD99}.
The intra-layer exchange interaction of Eq.~(\ref{X-intra}) scales linearly with
$r_s$, is always negative, and is important in favoring spin-polarized ground-state configurations. Regarding the inter-layer exchange contribution of 
Eq.~(\ref{X-inter}), its scaling with $r_s$ and $d^*$ is less trivial, mainly due to the presence of the ratio $d^*/r_s$ in the 
arguments of the exponentials. As discussed in the Appendix A, in the limit $d^*/r_s \ll 1$, and by expanding the 
exponentials this term acquires a leading linear dependence in $d^*$, like the Hartree electrostatic contribution. In the opposite limit $d^*/r_s \gg 1$, the exponential
terms are small and the inter-layer exchange scales linearly with $r_s$. According to our numerical evidence, this term is always positive and,
in consequence, its optimum configuration is either $\theta=0$ or the spin-balanced case $\eta_{a\sigma}=\eta_{b\sigma}=\eta_{c\sigma}$, as in both cases
$E_0^{\,\text{X-inter}}=0$.

\section{Results}
The total energy  in the 6-parameter space is plagued by  local  minima, that makes the task of
finding the global minimum a difficult numerical challenge. We also  have to minimize fulfilling the constraint 
$0\leq\eta_{i}\leq1$, with the minimum being just at the boundary in some cases. 
To carry out the minimization  we have partitioned the 6-parameter space in  (typically) $10^6$ regions. 
Starting from a central point for each region, we find a local  minimum using a Simplex algorithm~\cite{GSL}. 
Then we found the global  energy minimum as the minimum among all regions. That procedure is easy to parallelize, 
in particular  we have implemented it using MPI (Message Passing Interface) facilities~\cite{MPI}.

Before discussing the full numerical results, we find convenient to analyze some particular limits of the trilayer
system, which admits either analytical or semi-analytical solutions. Most of the ground-state configurations will appear 
already in these simple limits, helping the understanding of the general cases discussed later.
\subsection{$d^* = 0$ limit}
In the limit of small distance between layers, the expression for the ground-state energy simplifies
greatly,
\begin{eqnarray}
 \frac{E_0(d^*=0)}{2/r_s^2} &=& \sum_{\alpha \sigma} \left( \eta_{\alpha \sigma}^2-\frac{8\,r_s}{3\pi}\eta_{\alpha \sigma}^{3/2}\right) -
 \frac{t^* \: r_s^2}{\sqrt{2}} (\eta_a-\eta_c) \sin\theta \nn \\
 &+& \frac{\varepsilon_2^*\:r_s^2}{2} \left[ (\eta_a+\eta_c) \frac{\sin^2\theta}{2}+\eta_b \cos^2\theta \right] \;,
 \label{zero}
\end{eqnarray}
 having made the choice $\varepsilon_1^* = \varepsilon_3^* = 0$, $\varepsilon_2^* \ne 0$.
\afterpage{
\begin{figure}[h]
\begin{center}
\includegraphics[width=8.6cm]{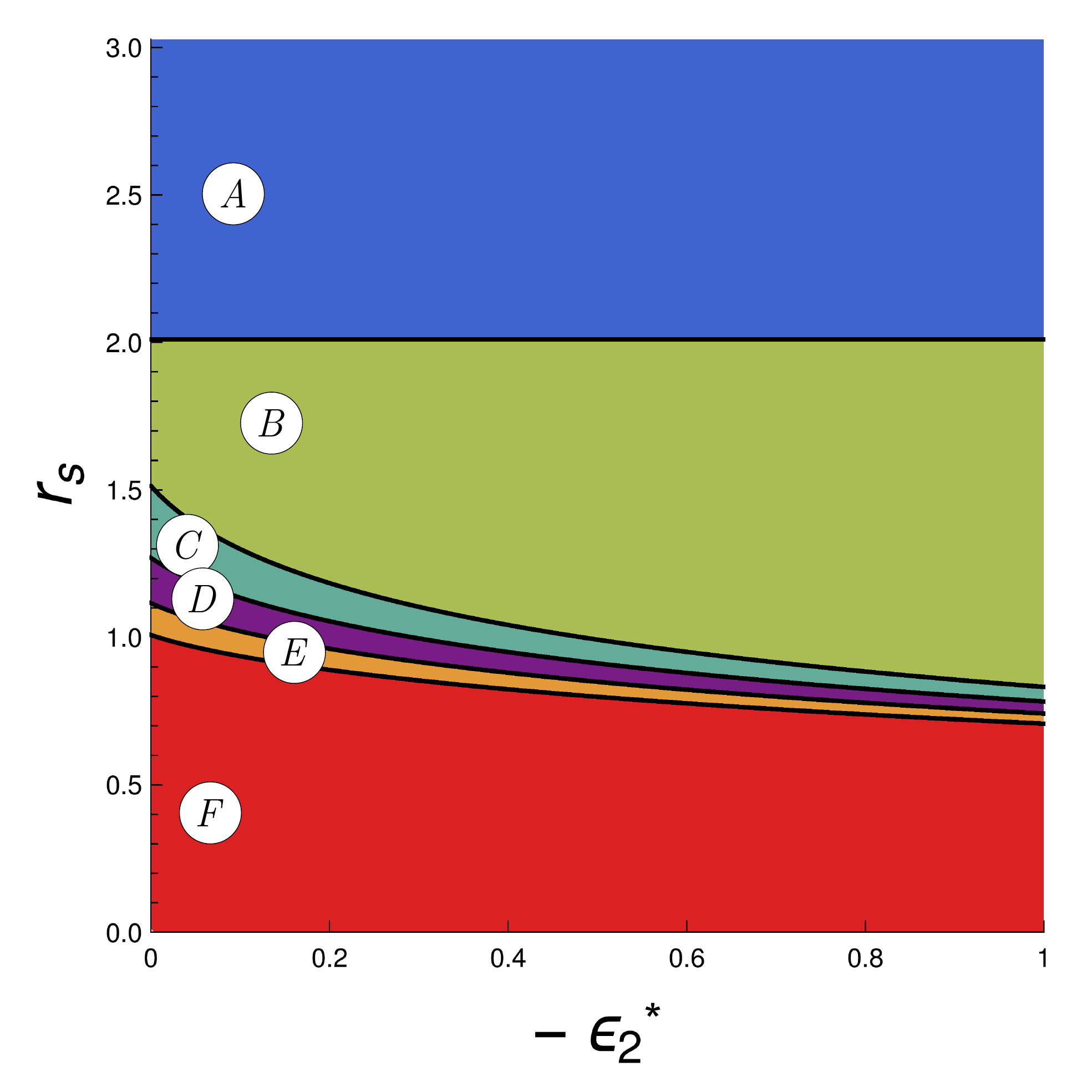}
\caption{\label{Fig2}Ground-state phase diagram, in the $r_s - \varepsilon_2^*$ plane, for $d^*=t^*=0$. 
The meaning of the different symbols is given in Table ~\ref{Table1}.}
\end{center}
\end{figure}
\begin{table}[h]
\begin{center}
\caption{Ground-state configurations for $d^*=t^*=0$. In all cases $\theta = 0$. 
Note that for $\phi=\theta=0$, the subband occupation factors are the same that the layer occupation factors.
 The symbol "sp" is used to identify spin-polarized states, while "ml", "bl", and "tl" represent monolayer, 
bilayer and trilayer configurations respectively. }
\label{Table1}
\begin{tabular*}{0.45\textwidth}{lcccccc}

\hline
Configuration & $\eta_{a\uparrow}$ & $\eta_{a\downarrow}$ & $\eta_{b\uparrow}$ & $\eta_{b\downarrow}$ &  $\eta_{c\uparrow}$ & $\eta_{c\downarrow}$ \\
\hline
\emph{A}: sp, ml & 0 & 0 & 1 & 0 & 0 & 0 \\
\emph{B}: ml & 0 & 0 & $\frac{1}{2}$ & $\frac{1}{2}$ & 0 & 0 \\
\emph{C}: bl & $1-2x$ & 0 & $x$ & $x$ & 0 & 0 \\
\emph{D}: bl & $\frac{1-2x}{2}$ & $\frac{1-2x}{2}$ & $x$ & $x$ & 0 & 0 \\
\emph{E}: tl & $y$ & $y$ & $x$ & $x$ & $1-2x-2y$ & 0 \\
\emph{F}: tl & $\frac{1-2x}{4}$ & $\frac{1-2x}{4}$ & $x$ & $x$ & $\frac{1-2x}{4}$ & $\frac{1-2x}{4}$ \\
\hline
\end{tabular*}
\end{center}
\end{table}
}
We display in Fig.~\ref{Fig2} the ground-state phase diagram which results from the numerical minimization
of Eq.~(\ref{zero}), for the case $t^*=0$~\cite{note1}.
The more prominent feature here is the growing of configuration \protect\emph{B} ($\eta_{b\uparrow}=\eta_{b\downarrow}=1/2$)
with respect to all configurations at lower values of $r_s$, as $\:-\:\varepsilon_2^*$ increases. This is clear physically:
as $\theta=0$, the configuration \protect\emph{B} is equivalent to $\eta_{2\uparrow}=\eta_{2\downarrow}=1/2$, and a
negative value of $\varepsilon_2^*$ favors the central well filling. The boundary between configurations \protect\emph{A} 
and
\protect\emph{B} does not depend on $\varepsilon_2^*$, since the energy of the term proportional to $\varepsilon_2^*$
is the same in both configurations (see Eq.~(\ref{zero})). The value $r_s=2.011$, which is the boundary between \protect\emph{A} and 
\protect\emph{B} is determined then by the balance between the intra-layer kinetic energy and the intra-layer exchange term.
Indeed, for $\varepsilon_2^*=0$ we can apply the analytical considerations from Ref.~[\onlinecite{HP00}], according
to which the critical density at which a transition occurs from a configuration with $p$ equally occupied components to another
configuration with $p+1$ equally occupied components is given by $r_s^{(0)}(p,p+1)=3\pi(1/\sqrt{p}+1/\sqrt{1+p})/8$. We obtain:
$r_s^{(0)}(1,2)\simeq 2.011$, $r_s^{(0)}(2,3)\simeq 1.513$, $r_s^{(0)}(3,4)\simeq 1.269$, $r_s^{(0)}(4,5)\simeq 1.116$,
and $r_s^{(0)}(5,6)\simeq 1.008$. These analytical results are exactly the transition points at $\varepsilon_2^*=0$ 
in Fig.~\ref{Fig2}, obtained numerically. 
For $t^*=0$ (the case analyzed in Fig.~\ref{Fig2}), exists the symmetry $\eta_a \leftrightarrow \eta_c$. This means, for example, that configuration \protect\emph{C},
corresponding to $\eta_{b\uparrow}=\eta_{b\downarrow}=x$, and $\eta_{a\uparrow}=1-2x$, is degenerate with the 
configurations $\eta_{a\downarrow}=1-2x$, or $\eta_{c\uparrow}=1-2x$, or $\eta_{c\downarrow}=1-2x$. For finite $t^*$,
we will see how some of these degeneracies are broken. Configurations \emph{A} and \emph{B} are the only
cases that we have found in the present work where the trilayer is actually a single-layer from the point of
view of the electronic distribution, with all electrons located in the central layer. The fully spin-polarized configuration \emph{A} is preferred over the 
unpolarized configuration \emph{B} at densities such that $r_s \ge 2.011$ by the action of the intra-subband exchange term in Eq.~(\ref{zero}),
which always favors spin-polarized configurations. As most of the transitions in this work, at the boundary between \emph{A} and \emph{B} configurations,
the occupation factors change abruptly.

\afterpage{
\begin{figure}[h]
\begin{center}
\includegraphics[width=8.6cm]{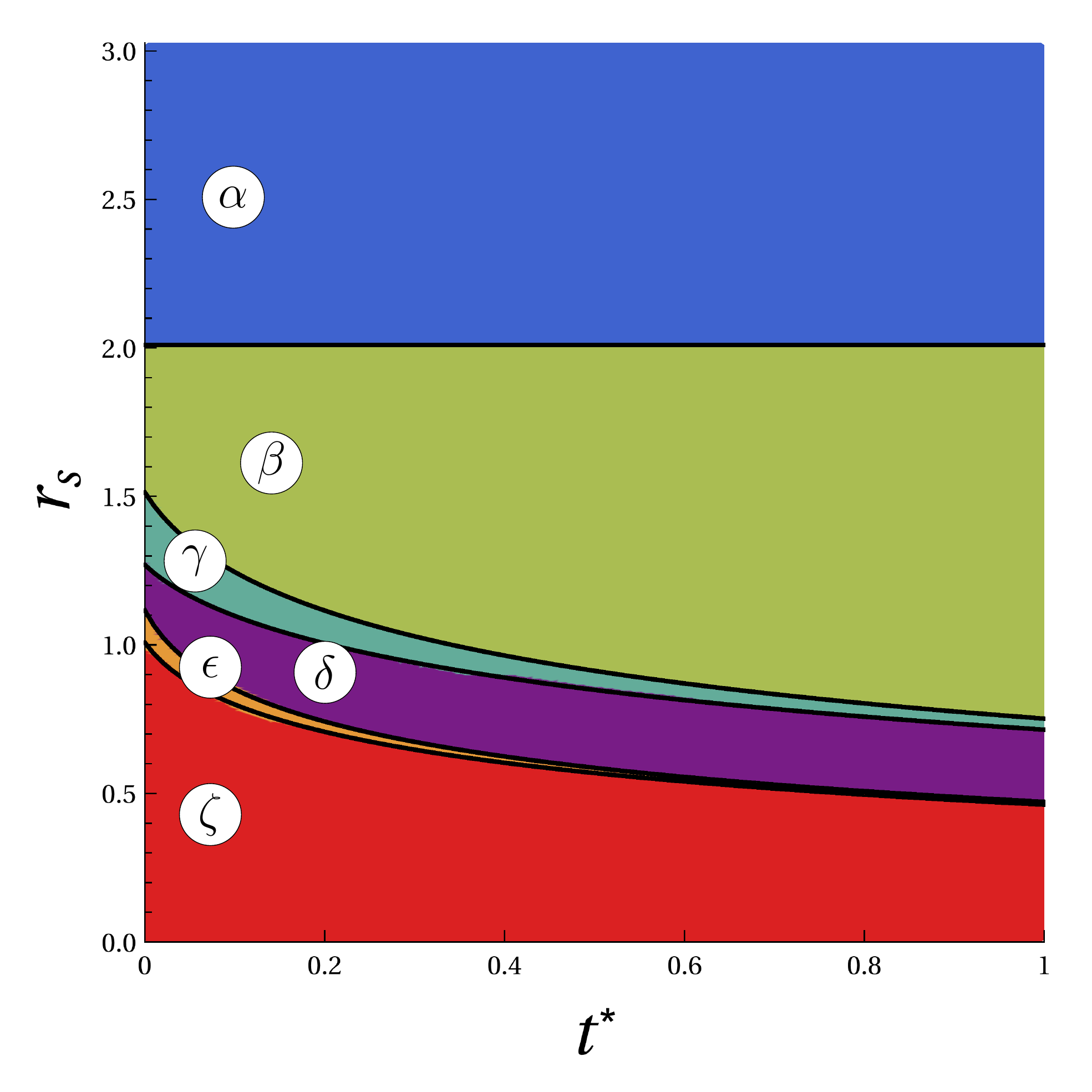}
\caption{\label{Fig3}Ground-state phase diagram, in the $r_s - t^*$ plane, for $d^*=\varepsilon_2^*=0$. 
The meaning of the different symbols is given in Table ~\ref{Table2}.}
\end{center}
\end{figure}
\begin{table}[h]
\begin{center}
\caption{Ground-state configurations for $d^*=\varepsilon_2^*=0$. In all cases $\theta = \pi/2$.}
\label{Table2}
\begin{tabular*}{0.45\textwidth}{lcccccc}

\hline
Configuration & $\eta_{a\uparrow}$ & $\eta_{a\downarrow}$ & $\eta_{b\uparrow}$ & $\eta_{b\downarrow}$ &  $\eta_{c\uparrow}$ & $\eta_{c\downarrow}$ \\
\hline
\emph{$\alpha$}: sp, tl & 1 & 0 & 0 & 0 & 0 & 0 \\
\emph{\bf $\beta$}: tl & $\frac{1}{2}$ & $\frac{1}{2}$ & 0 & 0 & 0 & 0 \\
\emph{\bf $\gamma$}: tl & $x$ & $x$ & $1-2x$ & 0 & 0 & 0 \\
\emph{\bf $\delta$}: tl & $x$  & $x$ & $\frac{1-2x}{2}$ & $\frac{1-2x}{2}$ & 0 & 0 \\
\emph{\bf $\epsilon$}: tl & $x$ & $x$ & $y$ & $y$ & $1-2x-2y$ & 0 \\
\emph{\bf $\zeta$}: tl & $x$ & $x$ & $y$ & $y$ & $\frac{1-2x-2y}{2}$ & $\frac{1-2x-2y}{2}$ \\
\hline
\end{tabular*}
\end{center}
\end{table}
}

As a way to understand the effect of the hopping parameter $t^*$, we display the 
ground-state configurations in the parameter space $r_s-t^*$, for $d^* = \varepsilon_2^* = 0$ in Fig.~\ref{Fig3}.
From Eq.~(\ref{zero}) with $\varepsilon_2^*=0$, the tunneling energy attains its minimum value with the choice
$\theta=\pi/2$, $\eta_a > \eta_c$. In the low-density limit $r_s \gg 1$, the tunneling term is the dominant
contribution and all the electrons are in the $a$-type subband (either in the $\alpha$ or $\beta$ configurations).
In the high-density limit $r_s \ll 1$, the physics is dominated by the intra-subband kinetic energy term, whose energy is
optimized (lower) by spreading electrons among the three subbands $a$, $b$, and $c$. The intra-subband exchange energy
contribution, scaling linearly with $r_s$ and favoring spin and pseudo-spin polarized states, plays an important
role at intermediate densities, by favoring spin-polarized configurations over spin-unpolarized
configurations as $r_s$ increases (i.e., $\beta \rightarrow \alpha$, $\delta \rightarrow \gamma$, $\zeta \rightarrow \epsilon$).
For increasing $t^*$, the configuration $\beta$ becomes more stable and gains area in the parameter space at the expense of the 
remaining configurations at
higher densities. In other words, at constant $r_s$, the occupancy of the bonding $a$-type subband increases
with $t^*$, for all $r_s < r_s^{(0)}(2,3)=1.513$, until the system falls into $\beta$ configuration. 

At first sight it is not clear why the ground-state configurations for $d^*=t^*=\varepsilon_2^*=0$ displayed
in Figs.~\ref{Fig2} and \ref{Fig3} are not the same. The point here is that in Fig.~\ref{Fig2} (\ref{Fig3})
we display the configurations which are the ground-states for {\em finite} values of $\varepsilon_2^*$ ($t^*$).
The case where the three external parameters $d^*$, $t^*$, and $\varepsilon_2^*$ strictly vanish is somehow
ill-defined, as the total energy then reduces to just the first two terms in Eq.~(\ref{zero}), that left the 
angle $\theta$ undetermined. Since any value of $\theta$ is allowed, the ground-state configuration is not
unique. The same consideration applies to the results displayed in Fig.~2 of 
Ref.~[\onlinecite{HP00}].

\begin{figure}[h]
\begin{center}
\includegraphics[width=8.6cm]{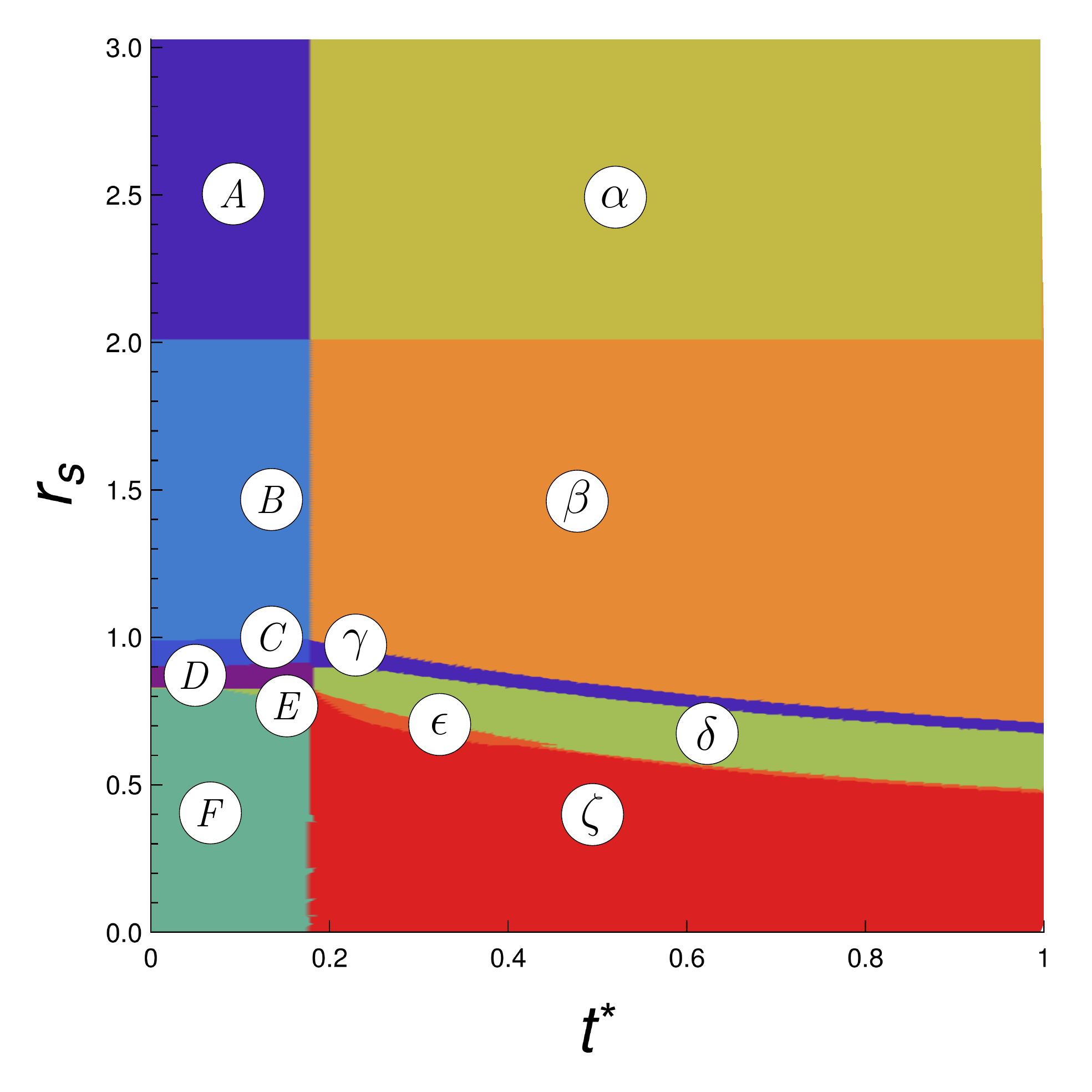}
\caption{\label{Fig4}Ground-state phase diagram, in the $r_s - t^*$ plane, for $d^*=0$, and 
$\varepsilon_2^* = - \: 0.5$. For $t^* \gtrsim 0.2$, all configurations are the same as in Fig.~\ref{Fig3}. For 
$0<t^* \lesssim 0.2$, the configurations \protect\emph{A}, \protect\emph{B}, and \protect\emph{F} are the same
as in Fig.~\ref{Fig2}. The three remaining configurations \protect\emph{C}, \protect\emph{D}, and \protect\emph{E} 
are similar
to the corresponding ones in Fig.~\ref{Fig2} regarding the subband occupancies, but with $\theta\ne0$.}
\end{center}
\end{figure}
The competition between the tunneling strength parameter $t^*$ and the central layer energy $\varepsilon_2^*$ is
shown in Fig.~\ref{Fig4}. For $0<t^* \lesssim 0.2$, the $\varepsilon_2^*$-driven configurations of Fig.~\ref{Fig2}
are the ground-state ones; for larger values of the tunneling parameter, the $t^*$-driven configurations of Fig.~\ref{Fig3} are
the ones with lower energies. The boundary between configurations \protect\emph{A} and \protect\emph{$\alpha$} may be
easily determined from Eq.~(\ref{zero}), and found to be $t^*=-\:\varepsilon_2^*/(2\sqrt{2})$. For $\varepsilon_2^*=-\:0.5$,
this gives $t^* \simeq 0.18$, in agreement with the numerically calculated boundary in Fig.~\ref{Fig4}. The boundary
does not depend on $r_s$, due to the equivalent occupation factors in configurations \protect\emph{A} (one fully occupied spin-polarized layer) and 
\protect\emph{$\alpha$} (one fully occupied spin-polarized subband), and the identical quadratic scaling with $r_s$ of both tunneling and $\varepsilon_s^*$ energy
contributions. For the same reason, the boundary between the \protect\emph{B} and \protect\emph{$\beta$} configurations
does not change with $r_s$. 

Considering configurations \protect\emph{C}, \protect\emph{D}, and \protect\emph{E}, they are similar to
the ones found in Fig.~\ref{Fig2} regarding the occupancies, but with $\theta \ne 0$. The minimizing angle $\theta$
may be found from Eq.~(\ref{zero}), as follows.
Calling $s=\sin\theta$, and given the simplicity of the $s$ dependence in the last terms, $E_0(d^*=0)$ may be optimized 
analytically with respect to $s$,
\begin{equation}
 \frac{dE_0(d^*=0)}{ds} = 0 \rightarrow s_{\text{opt}} = \frac{\sqrt{2}\:t^*(\eta_a-\eta_c)}{\varepsilon_2^*\:(\eta_a-2\eta_b+\eta_c)} \; .
\end{equation}
This is quite reasonable: for $t^* \ne 0$, and given that in configurations 
\protect\emph{C}, \protect\emph{D}, and \protect\emph{E}, $\eta_a \ne \eta_c$, the system gains some tunneling energy by
allowing that $\theta \ne 0$, although it is found numerically in all cases that the value of the minimizing angle is small.

It is worth of emphasize that at the (\protect\emph{A},\protect\emph{$\alpha$}) boundary crossing, 
the trilayer system suffers an abrupt reaccommodation 
of the electronic charge. This is easily seen from the following set of equations relating the occupation densities
in the layer and subband basis:
\begin{eqnarray}
 \eta_{1\sigma} &=& \cos^4\left(\frac{\theta}{2}\right)\eta_{a\sigma} + \frac{\sin^2\theta}{2} \: \eta_{b\sigma} + 
               \sin^4\left(\frac{\theta}{2}\right)\eta_{c\sigma} \; , \nn \\
 \eta_{2\sigma} &=& \frac{\sin^2\theta}{2} \: (\eta_{a\sigma}+\eta_{c\sigma}) + \cos^2\theta \: \eta_{b\sigma} \; , \nn \\            
 \eta_{3\sigma} &=& \sin^4\left(\frac{\theta}{2}\right)\eta_{a\sigma} + \frac{\sin^2\theta}{2} \: \eta_{b\sigma} + 
               \cos^4\left(\frac{\theta}{2}\right)\eta_{c\sigma} \; . \nn \\
\end{eqnarray}
Since in the \protect\emph{A} configuration $\eta_{b\uparrow}=1$ and $\theta=0$, this means in real space that
$\eta_{2\uparrow}=1$. In the \protect\emph{$\alpha$} configuration, since instead $\eta_{a\uparrow}=1$ and 
$\theta=\pi/2$, this translates in real space to the distribution $\eta_{1\uparrow}=\eta_{3\uparrow}=1/4$, 
$\eta_{2\uparrow}=1/2$. In words, at the (\protect\emph{A},\protect\emph{$\alpha$}) transition the system passes from a monolayer to a trilayer
configuration, as tunneling increases.
It is interesting to note that a ``balanced'' configuration in the $\{ 1,2,3 \}$ layer space is also a ``balanced'' 
configuration in the $\{ a,b,c \}$ subband space.

\subsection{Numerical results for the full model}
In Fig.~\ref{Fig5} we display the numerical results for the ground-state configurations, in the $r_s - d^*$
plane, with the layer distance dependent hopping $t^*(d^*)$, as given by Eq.~(\ref{tp2}) in Appendix B. 
For $d^* \ll 1$, the hopping is sizable ($t^*(0)\simeq 1 $), and the ground-state
configurations are the same as in Fig.~\ref{Fig3}. For $d^* \gg 1$, the Hartree term instead becomes predominant,
and the zero-tunneling and zero-site energy ground-state configurations of Ref.~[\onlinecite{HP00}] are obtained. The quadratic scaling of $E_0^{\text T}$ with
$r_s$, however, stabilizes the tunneling-driven configurations as $r_s$ increases. In other words, for enough large values of $r_s$,
the trilayer system always enter in a tunneling dominated regime.
The configuration labeled \emph{P2'}, which is different from the spontaneous inter-layer coherent state \emph{P1} configuration found previously 
($\eta_{b\uparrow}=1,\theta=\pi/2$)~\cite{HP00}, results from a competition between the hopping and inter-subband exchange contributions.
The \emph{P2'} configuration may be understood as a compromise between the tunneling stabilized $\beta$ configuration, and 
the Hartree induced \emph{P2} configuration. In passing from $\beta$ to \emph{P2}, $\theta$ vanishes and the system looses tunneling
energy. In \emph{P2'}, on the other side, the system keeps the gain in tunneling energy by allowing $\theta \ne 0$, but at
the same time minimizes to some extent the inter-subband exchange energy contribution by inducing a more ``balanced'' subband space configuration
(one subband is occupied in $\beta$, two subbands are occupied in \emph{P2'}).
\afterpage{
\begin{figure}[h]
\begin{center}
\includegraphics[width=8.6cm]{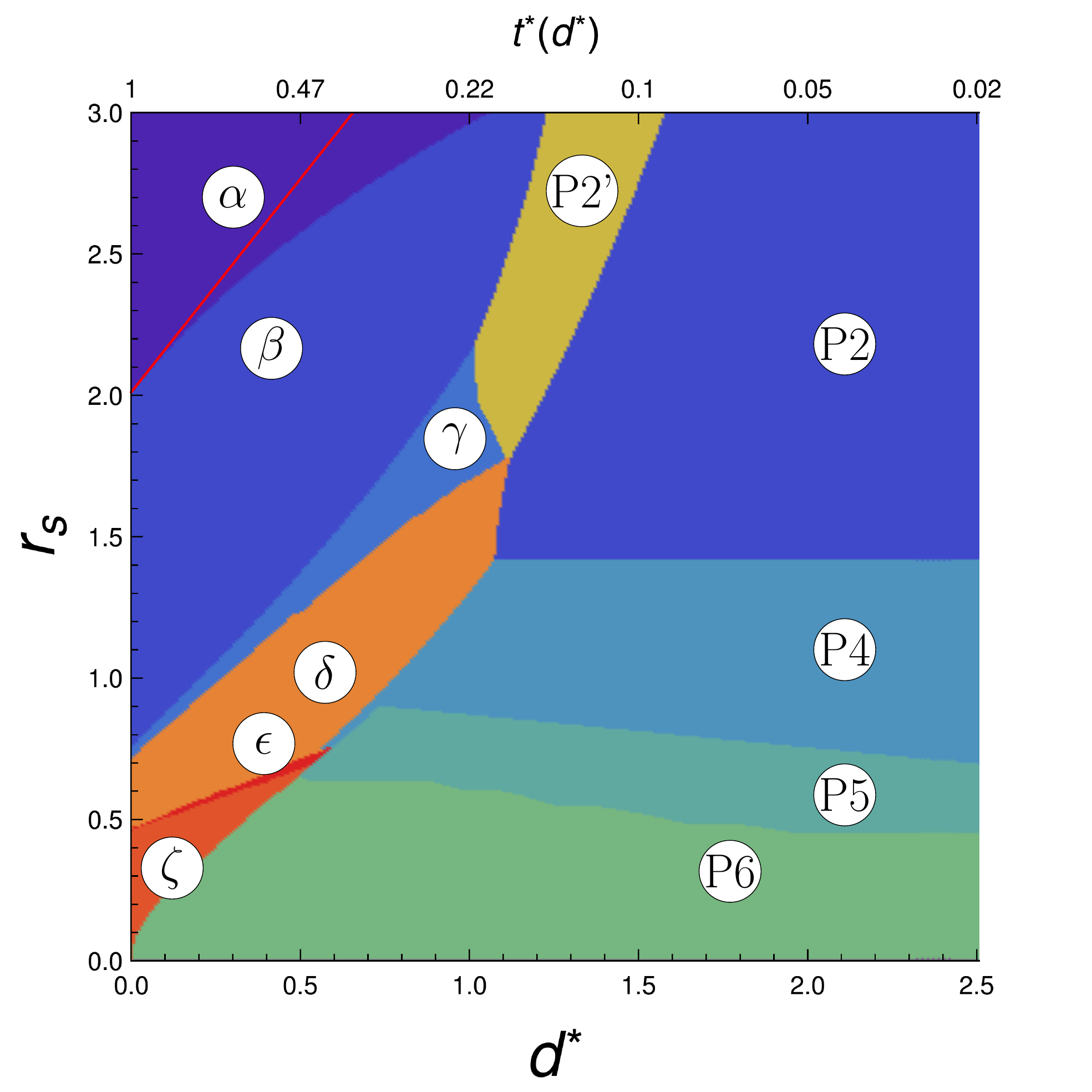}
\caption{\label{Fig5}Ground-state phase diagram, in the $r_s - d^*$ plane, for $\varepsilon_2^* = 0$ and
with the layer separation dependent hopping $t^*(d^*)$, as explained in the text. For small values of $d^*$, 
the ground-state configurations are the same of Fig.~\ref{Fig3}. For large values of $d^*$, the meaning of 
the different symbols is given in Table ~\ref{Table3}. The straight line corresponds to the approximation given 
in Eq.~(\ref{linear-boundary}) for the boundary between $\alpha$ and $\beta$ configurations.}
\end{center}
\end{figure}

\begin{table}[h]
\begin{center}
\caption{Ground-state configurations for $\varepsilon_2^* = 0$,
with the layer separation dependent hopping $t^*(d^*)$, and for large values of $d^*$. 
Configuration \emph{P3} is present between \emph{P4} and \emph{P5}, but not visible in the figure 
because it is very narrow. The symbol ``sp*'' is used to indicate that this configuration is 
degenerated with a non-polarized configuration obtained flipping the spin of a component.}
\label{Table3}\begin{tabular*}{0.42\textwidth}{lccccccc}

\hline
Configuration & $\eta_{a\uparrow}$ & $\eta_{a\downarrow}$ & $\eta_{b\uparrow}$ & $\eta_{b\downarrow}$ &  $\eta_{c\uparrow}$ & $\eta_{c\downarrow}$ &       $\theta$ \\
\hline
\emph{P2'}: sp, tl & $x$ & 0 & $1-x$ & 0 & 0 & 0 & $\pi/2$ \\
\emph{P2}: sp*, bl & $\frac{1}{2}$ & 0 & 0 & 0 & $\frac{1}{2}$ & 0 & 0 \\
\emph{P3}: bl & $x$  & $1-2x$ & 0 & 0 & $x$ & 0 & 0 \\
\emph{P4}: bl & $\frac{1}{4}$ & $\frac{1}{4}$ & 0 & 0 & $\frac{1}{4}$ & $\frac{1}{4}$ & 0 \\
\emph{P5}: tl & $x$  & $x$ & $1-4x$ & 0 & $x$ & $x$ & 0 \\
\emph{P6}: tl & $x$  & $x$ & $\frac{1-4x}{2}$ & $\frac{1-4x}{2}$ & $x$ & $x$ & 0 \\
\hline
\end{tabular*}
\end{center}
\end{table}
}

Another interesting feature of the phase diagram in the Fig.~\ref{Fig5} is the behavior of the frontier between the \emph{$\alpha$} 
(spin-polarized) and \emph{$\beta$} (non spin-polarized) configurations. 
For $d^*=0$, the limit is in the expected value of $r_s=2.011$ (as in Fig.~\ref{Fig3}). However, when $d^*$ grows and 
the associated $d^*$-dependent hopping $t^*(d^*)$ diminishes, 
the frontier between polarized and non-polarized configurations is moved to lower densities. 
Even for the lowest density considered (at $r_s=3.0$), it is possible to find a situation with the trilayer in a paramagnetic configuration. 
This is due to the $d^*$ dependence of the inter-subband exchange contribution. As explained in Appendix A, in the limit $d^*/r_s \rightarrow 0$,
one gets a linear dependence with $d^*$ for this term, as shown in Eq.~(\ref{A5}). 
Considering that the tunneling and Hartree energies are the same in \emph{$\alpha$} and \emph{$\beta$} configurations, the boundary between both can be found
by imposing the condition that the sum of the intra-subband kinetic energy, and the intra- and inter-subband exchange energies be the same in both cases.
In the limit $d^*/r_s \rightarrow 0$, one can use the expansion in Eq.~(\ref{A5}) for the inter-subband exchange energy and obtain the equation for the boundary 
\begin{equation}
 r_s^{\alpha \beta} (d^*)= \frac{3\pi}{8}\left(1+\frac{1}{\sqrt{2}} \right) \left( 1 + \frac{3d^*}{4} \right) \simeq 2.011 \left( 1 + \frac{3d^*}{4} \right) \: , 
\label{linear-boundary}
\end{equation}
valid to linear order in $d^*$. The important point here is that the gain in energy of the \emph{$\beta$} configuration comes from the inter-subband exchange interaction,
which in turn is a consequence that the tunneling between layers is finite, allowing a subband mixing angle $\theta \ne 0$.
For even higher $d^*$'s, the \emph{$\beta$} configuration turns into \emph{P2'}. 
When the layers are more separated, the angle $\theta$ flips from $\pi/2$ to 0, the system lost tunneling energy and configuration \emph{P2} becomes the one with the lowest energy.

\begin{figure}[h]
\begin{center}
\includegraphics[width=8.6cm]{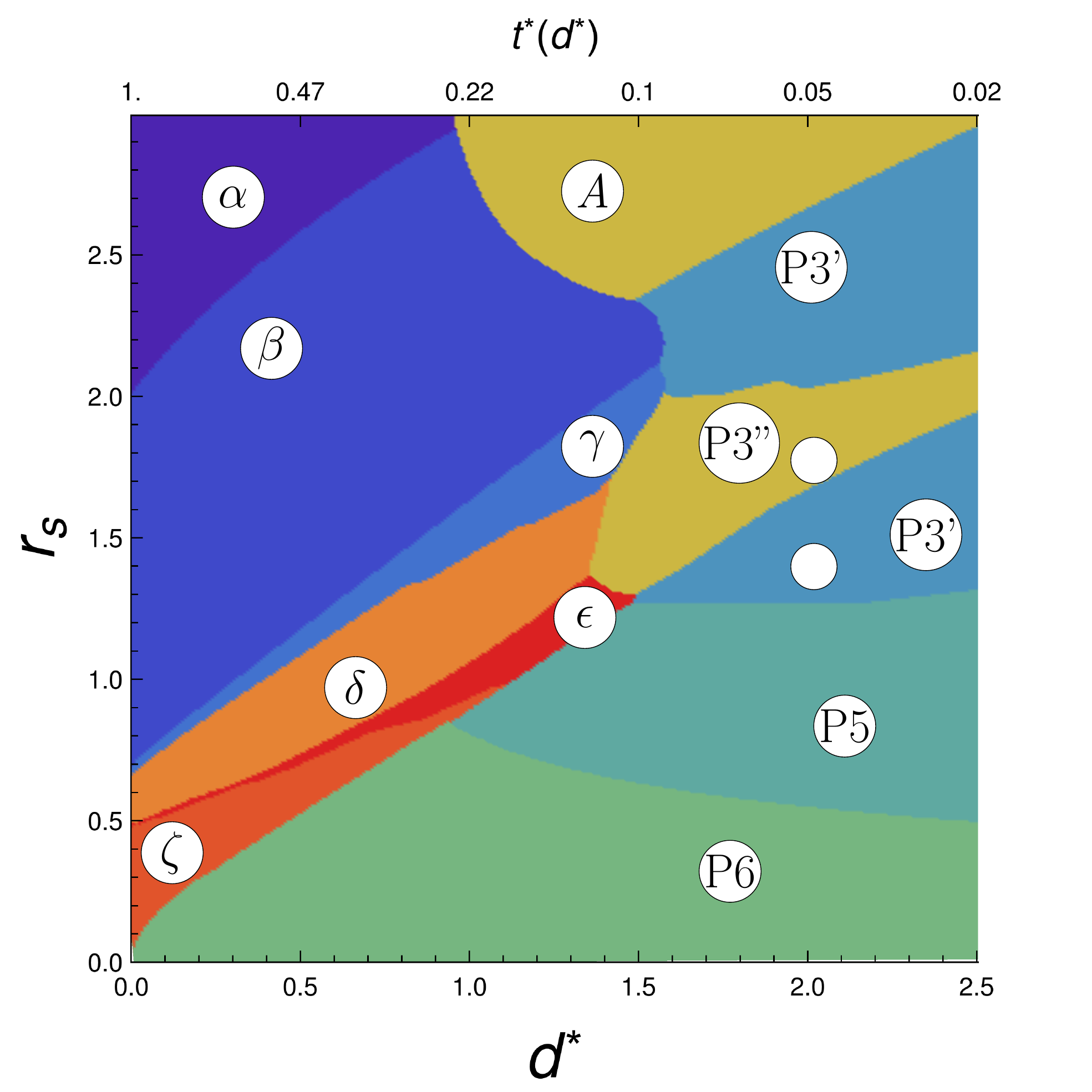}
\caption{\label{Fig6}Ground-state configurations in the parameter space $r_s-d^*$, for 
$\varepsilon_2^* = - \: 0.8$. The new configurations \protect\emph{P3'} and \protect\emph{P3''} are as follows:
\protect\emph{P3'},   $\eta_{a \uparrow }=\eta_{c \uparrow }= x$, $\eta_{b\uparrow}=1-2x,\theta=0$;
\protect\emph{P3''},  $\eta_{a \uparrow }= x, \: \eta_{b\uparrow }= y, \: \eta_{c\uparrow }= 1-x-y, \: \theta=\pi/2$.
The two empty circles correspond to the experimental sample of Ref.~[\onlinecite{SSS98}] (see text).}
\end{center}
\end{figure}
The only difference between Fig.~\ref{Fig5} and Fig.~\ref{Fig6} is that in the former $\varepsilon_2^*=0$, while
in the latter $\varepsilon_2^*=-\:0.8$. The main consequences are the disappearance of configurations 
\protect\emph{P2} and \protect\emph{P4} of Fig.~\ref{Fig5} in Fig.~\ref{Fig6}, being replaced by configurations
\protect\emph{P3'} and \protect\emph{P3''}, and the growing stability of the tunneling-driven configurations on
the left-hand side of the diagram. This last feature is easy to understand: by doing $\varepsilon_2^* < 0$ the 
occupation of the central layer increases, and this favors the stability of the \protect\emph{$\alpha$} and
\protect\emph{$\beta$} configurations particularly, that when translated to layer space have a central layer with
twice the occupancy of the two side layers. This effect is also reflected in configurations 
\emph{$\gamma$}, \emph{$\delta$}, \emph{$\epsilon$}, \emph{$\zeta$}, although somehow
in a less strength, due to the decreasing influence of tunneling and layer energies for decreasing $r_s$. 

Regarding the disappearance of the \emph{P2} and \emph{P4} configurations which are present in 
Fig.~\ref{Fig5} but not in Fig.~\ref{Fig6}, this is due to the fact that both configurations have zero occupancy
of the central layer, and this is in direct conflict with the fact that $\varepsilon_2^* < 0$. The new stable 
configurations \emph{P3'} and \emph{P3''}, on the other side, have a finite occupancy of the central
layer.

As in the $\varepsilon^*_2 = 0$ case, the non spin-polarized configuration $\beta$ is still present even at low densities $(r_s > 2.0)$ and 
low and medium inter-layer distances. However, when $d^*$ increases (and consequently, 
the hopping decreases) the configuration \emph{A} is stabilized. This configuration implies that all the electrons are in the central 
layer with the same spin projection. It is stabilized by the exponential decay of the tunneling parameter with $d^*$, that somehow recreates
the situation in Fig.~\ref{Fig2}, in the limit of low densities. For larger $d^*$, 
the Hartree energy becomes important and a situation with the three layers occupied, like \emph{P3'}, is preferred.

\begin{figure}[h]
\begin{center}
\includegraphics[width=8.6cm]{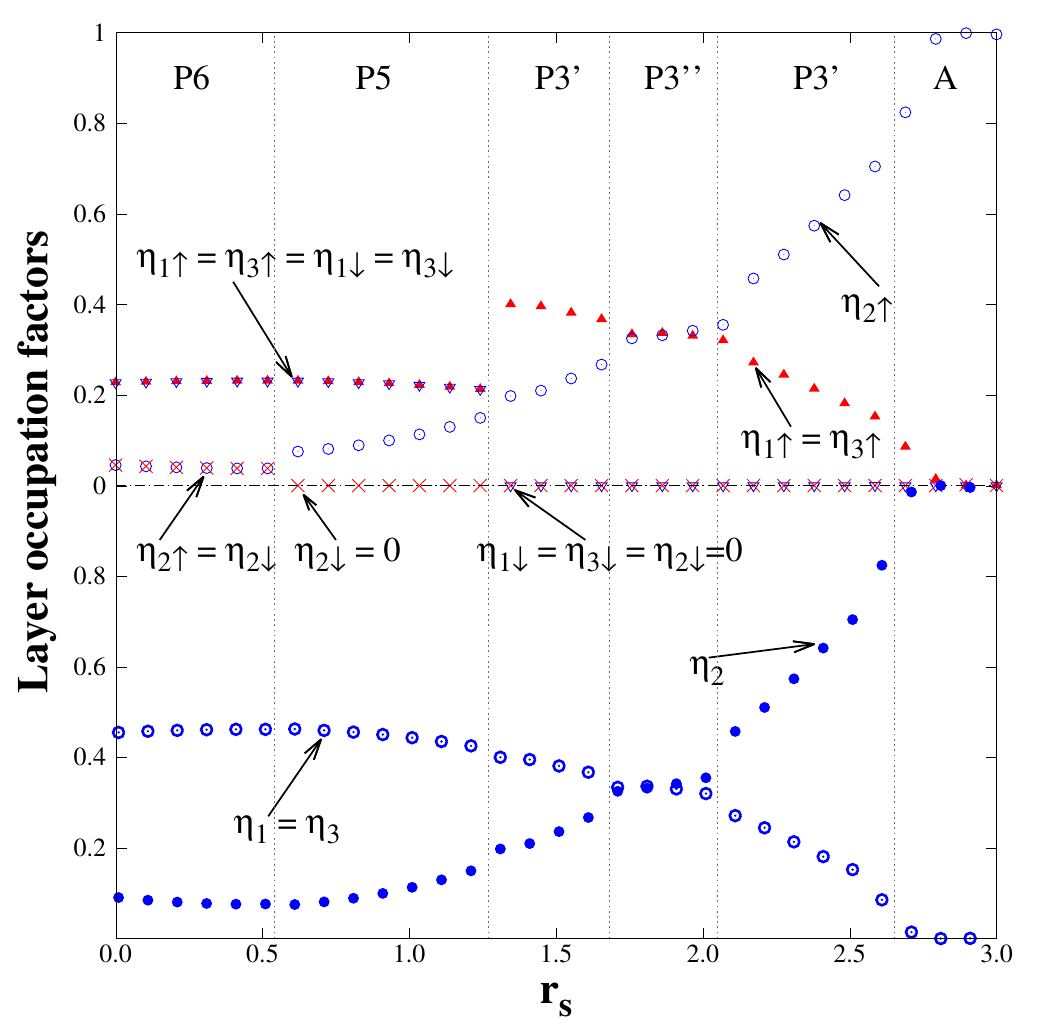}
\caption{\label{Fig7}Layer occupation factors for a trilayer with $d^*=2$, $\varepsilon_2^*=-\:0.8$ and 
$t^*(d^*=2)=0.05$. In the upper (lower) panel spin discriminated (undiscriminated) occupation factors are shown.
At the tiny dashed vertical lines, there is a change in the ground-state configuration.}
\end{center}
\end{figure}

Finally, we display in Fig.~\ref{Fig7} the layer occupancies as function of $r_s$, for fixed $d^* \; (\simeq 2)$,
$t^* \; (\simeq 0.05)$ and $\varepsilon_2^* \; (\simeq -\:0.8)$~\cite{SSS98}.
For $r_s \rightarrow 0$, the layer occupancies in the figure are well reproduced by the analytical estimates
given in Ref.~[\onlinecite{HP00}], according to which the layer occupation factors in the \emph{P6} configuration are given by
\begin{equation}
 \eta_a = \eta_c = \frac{1}{3}\left( 1+\frac{\xi}{2} \right) \quad \quad \eta_b = \frac{1}{3}\left( 1-\xi \right) \; ,
\end{equation}
with $\xi(d^*)=(1+3/4d^*)^{-1}$. Evaluating these expressions at $d^*=2$, we obtain that $\eta_a=\eta_c=5/11 \simeq 0.45$,
$\eta_b=1/11 \simeq 0.09$, in good agreement with the numerical values at $r_s=0$ in Fig.~\ref{Fig7}. To obtain  these
analytical estimates, only the intra-layer kinetic and Hartree energies contributions to the total energy are
considered, as dictated by the scaling behavior of the total energy in the high-density limit.
This abrupt redistribution of charge among the layers is reminiscent of the abrupt charge transfer between the 
ground- and first-excited subbands of a single semiconductor quantum well, observed experimentally in 
Ref.~[\onlinecite{GHTERPG02}] and
discussed theoretically in Ref.~[\onlinecite{RP07}].

For increasing $r_s$, the sequence of transitions is as follows: 
{\emph{P6}} $\rightarrow$ {\emph{P5}} $\rightarrow$ \emph{P3'} $\rightarrow$ \emph{P3''} $\rightarrow$ \emph{P3'} $\rightarrow$ {\emph{A}}.
The transitions \emph{P6} $\rightarrow$ \emph{P5}, \emph{P5} $\rightarrow$ \emph{P3'}, and \emph{P3'} $\rightarrow$ \emph{A} are clearly discontinuous,
as at each one of them a given layer occupation factor abruptly passes from a finite value to zero. The interesting 
``reentrant'' sequence \emph{P3'} $\rightarrow$ \emph{P3''} $\rightarrow$ \emph{P3'}, on the other side, involves only smooth occupation
changes at the boundaries. The full sequence can be understood as a consequence of the progressive filling of the 
central well as $r_s$ increases. The quadratic scaling with $r_s$ of the term proportional to $\varepsilon_2^*$ in 
Eq.~(\ref{K}) makes this term to become dominant for large $r_s$, culminating with the stabilization of the \emph{A}
configuration, where all electrons are in the central layer, and fully spin polarized.
The ``re-entrant'' sequence \emph{P3'} $\rightarrow$ \emph{P3''} $\rightarrow$ \emph{P3'} is interesting, since it reflects once more the 
importance of the inter-layer exchange energy contribution $E_0^{\text{X-inter}}$, and its competition with the tunneling
energy $E_0^{\text T}$. In the \emph{P3'} configuration, $\theta=0$, as this is one of the two possible ways to cancel
the positive contribution of $E_0^{\text{X-inter}}$, although that leads to a vanishing of the gain in tunneling
energy too. But for $1.68 \lesssim r_s \lesssim 2.05$, the \emph{P3''} ``balanced'' configuration becomes the one
with the lowest energy. Here, since $\eta_{1\uparrow} \simeq \eta_{2\uparrow} \simeq \eta_{3\uparrow}$, 
$E_0^{\text{X-inter}}$ is minimized using the second option available for making it as small as possible: equal
occupancy of all the subbands. This allows that $\theta \neq 0$, and leads to a gain of the tunneling energy.
As $r_s$ increases further, the equal occupancy constraint cannot be maintained any more, and the system returns to the 
unbalanced \emph{P3'} ground-state configuration,  but now with a preferential occupancy of the central layer.

It should be noted that the three parameters $d^*$, $t^*$, and $\varepsilon_2^*$ used in Fig.~\ref{Fig7}
were directly obtained from the triple semiconductor quantum well system studied experimentally in Ref.~[\onlinecite{SSS98}],
as explained in Appendix B. 
After adjustment of these parameters, our model reproduces qualitatively the main features of the more
elaborated calculations reported in Ref.~[\onlinecite{SSS98}], using density-functional-theory in the Local
Density Approximation. Due to the associated computational cost, these types of calculations are usually restricted
to a given particular set of parameters. In particular, it is reported in Ref.~[\onlinecite{SSS98}] that
$\eta_1 \simeq \eta_2 \simeq \eta_3 \simeq 1/3$ at $N \simeq 10.8 \times 10^{10} \text{ cm}^{-2}$ ($r_s \simeq 1.74$, 
corresponding to the upper empty white circle in Fig.~\ref{Fig6}), and that
$\eta_1/\eta_2 \simeq \eta_3/\eta_2\simeq 2$ at $N \simeq 16.7\times 10^{10} \text{ cm}^{-2}$ ($r_s \simeq 1.4$, 
corresponding to the lower empty white circle in Fig.~\ref{Fig6})~\cite{note2}. 
These LDA determined layer occupation factors are in good qualitative agreement with the ones from our model, 
as shown in Fig.~\ref{Fig7}. This gives us some confidence that our simple model is able to reproduce the results
of more elaborate calculations, and that it may be useful for the design of real samples and for understanding the 
physics of trilayer semiconductor systems.

One may wonder how much of the findings of the present work can be extrapolated to other apparently similar
2DEG's, as for example those formed at the interface between two band insulators, such as 
$SrTiO_3$ and $LaAlO_3$~\cite{MBHMT08}.
These are fascinating systems, showing evidence of superconducting and ferromagnetic properties, 
simultaneously~\cite{B11}. 
Unfortunately, our simple theoretical model is not suitable for capturing the main physics needed for describing
these 2DEG's at oxide interfaces. In the first place, the experimental results suggest the presence of two type of
carriers at the interface~\cite{S09}: one type is mobile, and presumably responsible of the superconducting features, and the 
other type of carriers seems to be localized, and responsible of the magnetic (ferromagnetic) features. Our model
has only mobile carriers, which are the ones that gives all the possible ground states presented above.
Other important difference is that we have assumed from the very outset that our semiconductor-based 2DEG has
translational invariance in the $x-y$ plane, an assumption that seems to be not justified in the case of the 
2DEG at oxide interfaces~\cite{B11}. And lastly, all the evidence in this latter case points to the importance of correlation
effects in describing their physics~\cite{HIKKNT12}, beyond the reach of our variational HF approximation, 
which on the other side
is reasonable for the treatment of the weakly-correlated semiconductor systems discussed here.

\section{Conclusions}
The possible ground-state configurations of a trilayer system have been determined, within the framework of a
variational Hartree-Fock approximation. The metallic layers are Coulomb coupled through the inter-layer Hartree
and exchange interactions, and also due to the tunneling between the neighboring layers. At high-density the 
system becomes quite simple, as the only remaining terms in the total energy in this regime are the intra-layer
kinetic energy, and the Hartree classical contribution. The low-density regime is dominated by two single-particle
effects introduced in this work: the tunneling between layers and the site energies. The inter-layer exchange
interaction is found to play an important role, helping in the stabilization of the balanced configuration,
in whose vicinity most of the experimental samples are designed. It is expected that the results presented in this
work for a wide range of parameters, may serve as a qualitative guide for the design of experimental samples,
and also be useful for the understanding of the physics of trilayer semiconductor systems at zero magnetic field.

\begin{acknowledgments}
DM acknowledges the support of ANPCyT under Grant PICT-2012-0379. CRP and PGB thank Consejo Nacional de Investigaciones 
Cient\'ificas y T\'ecnicas (CONICET) for partial financial support and ANPCyT under Grant PICT-2012-0379. 
We acknowledge Shinichi Amaha for motivate us to perform the present study and for many stimulating discussions.
\end{acknowledgments}

\appendix
\section{Some analytical results for the inter-exchange energy contribution}
Considering that $E_0^{\text{X-inter}}$ is the only contribution to the total energy whose dependence with
$r_s$ and $d^*$ is not fully explicit from its definition in Eq.~(\ref{X-inter}), we find convenient to analyze here
some limits for it, where analytical results are available.
In first place, the integrals $I_{\alpha \beta \sigma}(q)$ in Eq.~(\ref{X-inter}) are given by
\begin{widetext}
\begin{eqnarray}
 I_{\alpha \beta \sigma}(q)&=&\eta_{\alpha \sigma}\Theta(\sqrt{\eta_{\beta \sigma}}-\sqrt{\eta_{\alpha \sigma}}-q) +
                       \eta_{\beta \sigma}\Theta(\sqrt{\eta_{\alpha \sigma}}-\sqrt{\eta_{\beta \sigma}}-q)                      
                       + \pi^{-1}\Theta(\sqrt{\eta_{\alpha \sigma}}+\sqrt{\eta_{\beta \sigma}}-q)
                       \Theta(q-\lvert\sqrt{\eta_{\alpha \sigma}}-\sqrt{\eta_{\beta \sigma}}\rvert) \nn \\                    
                       &\times&\left\{\eta_{\alpha \sigma}\left[\cos^{-1} \left(\frac{q+K_0}{2\sqrt{\eta_{\alpha \sigma}}}\right) 
 -\left(\frac{q+K_0}{2\sqrt{\eta_{\alpha \sigma}}}\right)\sqrt{1-\left(\frac{q+K_0}{2\sqrt{\eta_{\alpha \sigma}}}\right)^2}\right]\right.
 \nn \\
 &+& \left.\eta_{\beta \sigma}\left[\cos^{-1} \left(\frac{q-K_0}{2\sqrt{\eta_{\beta \sigma}}}\right) 
 -\left(\frac{q-K_0}{2\sqrt{\eta_{\beta \sigma}}}\right)\sqrt{1-\left(\frac{q-K_0}{2\sqrt{\eta_{\beta \sigma}}}\right)^2}\right]
 \right\} \; , \nn \\
\end{eqnarray}
\end{widetext}
and $I_{\alpha \beta \sigma}(q) = I_{\beta \alpha \sigma}(q)$.
Here, $K_0=(\bar{k}_{\alpha \sigma}^2-\bar{k}_{\beta \sigma}^2)/q$, with $\bar{k}_{\alpha \sigma}=k_{\alpha \sigma}/k_F$, 
and $k_F=\sqrt{4 \pi N}$.
For $\alpha=\beta$, the expression simplifies to 
 \begin{eqnarray}
 I_{\alpha \alpha \sigma}(q) &=\frac{2\eta_{\alpha \sigma}}{\pi} \Theta(2\sqrt{\eta_{\alpha \sigma}}-q) \left[\vphantom{\sqrt{1-\left(\frac{q}{2\sqrt{\eta_{\alpha \sigma}}}\right)^2}}\cos^{-1} \left(\frac{q}{2\sqrt{\eta_{\alpha \sigma}}}\right)\right.\nn  \\
  &\quad \left. {} -\left(\frac{q}{2\sqrt{\eta_{\alpha \sigma}}}\right)\sqrt{1-\left(\frac{q}{2\sqrt{\eta_{\alpha \sigma}}}\right)^2}\right].
\end{eqnarray}
In the limit $d^*/r_s\rightarrow 0$, and to the first order in $d^*/r_s$, the term $E_0^{\,\text{X-inter}}$ 
can be calculated explicitly:
\begin{eqnarray}
 E_0^{\,\text{X-inter}} \left(\frac{d^*}{r_s} \rightarrow 0 \right) &\simeq& 2 d^*\sin^2\theta \sum_{\sigma} \int\limits_0^{\infty} dq\:q \nn \\
 &\times& \left(I_{aa\sigma}+2I_{bb\sigma}+I_{cc\sigma}-2I_{ab\sigma}-2I_{cb\sigma} \right) \nn \\
                              &-& \frac{d^* \sin^4\theta}{2} \sum_{\sigma} \int\limits_0^{\infty} dq\:q \left( I_{aa\sigma}+4I_{bb\sigma}\right. \nn \\
                              &+& \left.I_{cc\sigma}+2I_{ac\sigma}-4I_{ab\sigma}-4I_{cb\sigma}\right). \nn \\
                              \label{X-inter-d-0} \;                                                             
\end{eqnarray}
Using the relation~\cite{HHDV00}
\begin{equation}
 2 \int\limits_0^{\infty}\left(I_{\alpha \alpha \sigma}+I_{\beta \beta \sigma}-2I_{\alpha \beta \sigma}\right)dq\:q=
 \left(\eta_{\alpha \sigma}-\eta_{\beta \sigma}\right)^2,
\end{equation}
it follows from Eq.~(\ref{X-inter-d-0}) that
\begin{eqnarray}
 E_0^{\,\text{X-inter}}\left( \frac{d^*}{r_s} \rightarrow 0 \right) &\simeq&d^{*}\sin^{2}\theta\left(1-\frac{\sin^{2}\theta}{2}\right) \nn \\
 &\times& \sum_{\sigma}\left[\left(\eta_{a\sigma}-\eta_{b\sigma}\right)^{2}+\left(\eta_{c\sigma}-\eta_{b\sigma}\right)^{2}\right] \nn \\
 &+& \frac{d^{*} \sin^{4}\theta}{4}\sum_{\sigma}\left(\eta_{a\sigma}-\eta_{c\sigma}\right)^{2}. 
 \label{A5}
\end{eqnarray}
From the above expression is easy to see that since in this limit $E_0^{\,\text{X-inter}}$ is always positive,
it is minimized either for $\theta=0$ or when all subbands are equally occupied. Assuming the case $\theta \neq 0$,
the equally occupied situation may be realized in two different ways: $i)$ spin-polarized, with 
$\eta_{a\uparrow}=\eta_{b\uparrow}=\eta_{c\uparrow}=1/3$ and all spin-down occupancies equal to zero, and
$ii)$ spin-unpolarized situation with all subband occupancies for both spins equally occupied and having the value
1/6. Owing to the presence of the remaining contributions to the total energy, the spin-polarized situation
is more stable in the low-density limit, while the spin-unpolarized situation is preferred in the high-density limit.

\section{Derivation of the tunneling parameter $t^*(d^*)$}
In a tight-binding like approximation, the hopping parameter between two neighboring 
semiconductor quantum wells separated by a distance $d$ may be estimated from the expression~\cite{Bastard}
\begin{eqnarray}
 t(d) = \int \phi_{loc}(z-d)\:V_b(z)\:\phi_{loc}(z) \: dz \; .
 \label{tp}
\end{eqnarray}
Here, $\phi_{loc}(z-d)$ and $\phi_{loc}(z)$ are the envelope normalized wavefunctions corresponding to the left and right
quantum wells, respectively, and $V_b(z)$ is the potential barrier between the two wells. 
Consistently with our trilayer model in Fig.~1, we have approximated each quantum well by an isolated
attractive delta-potential of strength $\alpha$. 
Within this model, $\phi_{loc}(z)= \sqrt{\alpha^*/(2a_0^*)}\exp{(-\alpha^*|z|/2a_0^*)}$, 
$V_b(z)=-\alpha\big[\delta(z)+\delta(z-d)\big]$, with $\alpha$ having
units of energy times length, and $\alpha^*=\alpha/(a_0^* \text{Ry}^*)$.
Replacing in Eq.~(\ref{tp}), and imposing the constraint that $t^*(0) \simeq 1$ (a reasonable physical choice) we obtain 
\begin{equation}
 t^*(d^*)=\exp{(-\frac{\alpha^*d^*}{2})} \; .
 \label{tp2}
\end{equation}
The free parameter $\alpha^*$ can be fixed now by imposing the second condition that for a 
given value of $d^*$, the hopping parameter should be equal to some convenient value, usually taken from a more elaborate
calculation. Solving equation above for $\alpha^*$, it yields
\begin{equation}
 \alpha^* = - \frac{2\ln(t^*)}{d^*} \; .
 \label{tp3}
\end{equation}
As an example of the use of these equations, using Eq.~(3.6) in Ref.~[\onlinecite{HMcD96}], the value of $t^*$
may be estimated from the electronic subband structure of a particular trilayer. For instance, using the 
Local-Density-Approximation theoretical results in Ref.~[\onlinecite{SSS98}] corresponding to a trilayer with
$d^* \simeq 2$, we obtain that $t^* \simeq 0.05$. Replacing in Eq.~(\ref{tp3}), $\alpha^* \simeq 2.97$. This gives
us the dependence of the hopping parameter with the distance between layers employed in Fig.~(\ref{Fig6}). 
The parameter $\varepsilon_2^*$ was estimated by following the considerations of Hanna and MacDonald in Ref.~[\onlinecite{HMcD96}],
for the same trilayer system.


\end{document}